\documentclass[aps,prl,twocolumn,showpacs,amsmath,amssymb,superscriptaddress,letterpaper]{revtex4}
\usepackage{times}
\usepackage{amsfonts}
\usepackage{mathrsfs}
\usepackage{mathtools}
\usepackage{graphicx,epstopdf}
\usepackage{dcolumn}
\usepackage{bm}
\usepackage{color}
\usepackage[colorlinks,bookmarks=false,citecolor=blue,linkcolor=red,urlcolor=blue]{hyperref}
\usepackage{booktabs}

\graphicspath{{figures/}}

\begin{document}

\title{Weyl and nodal ring magnons in three-dimensional honeycomb lattices}
\author{Kangkang Li}
\affiliation{Beijing National Laboratory for Condensed Matter Physics, and Institute of Physics, Chinese Academy of Sciences, Beijing 100190, China}
\affiliation{School of Physics, University of Chinese Academy of Sciences,  Beijing 100049, China}

\author{Jiangping Hu}
\affiliation{Beijing National Laboratory for Condensed Matter Physics, and Institute of Physics, Chinese Academy of Sciences, Beijing 100190, China}
\affiliation{School of Physics, University of Chinese Academy of Sciences,  Beijing 100049, China}
\affiliation{Collaborative Innovation Center of Quantum Matter, Beijing, China}

\begin{abstract}
We study the topological properties of magnon excitations in a wide class of three dimensional (3D) honeycomb lattices with
ferromagnetic ground states. It is  found that they host nodal ring magnon excitations.   These rings locate
on the same plane  in the momentum space. The rings can be gapped  by Dzyaloshinskii-Moriya (DM) interactions to form two Weyl points with opposite charges.
 We explicitly discuss these physics
in the simplest 3D honeycomb lattice, the hyperhoneycomb lattice and show  drumhead and arc surface states in the nodal ring and Weyl phases, respectively,  due to
the bulk-boundary correspondence.
\end{abstract}

\maketitle

\paragraph{Introduction.} Two dimensional (2D) honeycomb lattice can be realized in graphene, silicene and many other
related materials. Its two Dirac points in the first Brillouin zone make it one of the most fascinating
research field in condensed matter physics. When the spin-orbit interaction (SOI)  is  large,
 the band structure of  the system is topologically nontrivial  and becomes a topological insulator \cite{Kane1,Kane2}.

 Naturally, there are also 3D honeycomb lattices \cite{JTWang,Taka,Bif,Modic,KinHo1,KinHo2,Mullen,Nasu1,Nasu2,Kimchi,SBLee,Herma}, and  their exotic properties have been explored, such as nodal line semimetal in body-centered orthorhombic $C_{16}$ \cite{JTWang},
loop-nodal semimetal \cite{KinHo1,Mullen} and topological insulator \cite{KinHo1} in the hyperhoneycomb lattice,
nodal ring \cite{Robert,Herma} and Weyl \cite{Herma} spinons of interacting spin systems in the hyperhoneycomb and stripy-honeycomb lattices,
and loop Fermi surface in other similar systems \cite{Mandal,Phillips,LSXie,RYu,Kim,Yama,HWeng}. Based on these studies,
Ezawa \cite{Ezawa} proposed a wide class of 3D honeycomb lattices constructed by two building blocks. These 3D
honeycomb lattices can display all loop-nodal semimetals which can  be gapped to be strong topological insulators  by SOI or point nodal semimetals
by SOI together with an antiferromagnetic order.

Recently, the topology of band has been extended to magnetic excitations as well \cite{Molina,Lifa,Shindou,Kwon,Owerre1,Owerre2,Feiye,Mook1,Mook2,Kangkang},   including magnon Chern insulators \cite{Molina,Lifa,Shindou,Kwon,Owerre1,Owerre2}, Weyl magnons \cite{Feiye,Mook1}, magnon nodal-line semimetals \cite{Mook2} and Dirac magnons \cite{Kangkang}. Interestingly, the magnon excitations on a 2D honeycomb lattice with a ferromagnetic ground state
 have two Dirac points
in the first Brillouin zone \cite{Owerre1}, and a proper DM interaction can gap the system into a magnon Chern insulator, reminiscent of the role of SOI  in  the graphene\cite{Kane1,Kane2}. Therefore, we can ask a natural question whether there exist  magnon excitations on 3D honeycomb lattices with special properties such as nodal lines  and nodal points,  similar to those in   electronic systems \cite{Ezawa}.

In this paper, we provide an affirmative answer to the above question. Use the linear spin-wave approximation, we study the magnon excitations on the wide class of 3D honeycomb lattices proposed in \cite{Ezawa}. We consider a Heisenberg model with isotropic nearest neighbor ferromagnetic exchange interactions.  We find that all the 3D honeycomb lattices host magnon nodal rings located in the $k_z=0$ plane. Furthermore, when the DM interaction is considered, every ring  can be gapped into two Weyl points with opposite charges. The position of the rings can be analytically obtained.
For a concrete example, we calculate the surface states of the simplest 3D honeycomb lattice, the hyperhoneycomb lattice. Due to
the bulk-boundary correspondence \cite{Hat1,Hat2}, there are drumhead and arc surface states in the nodal ring and Weyl phases respectively.

\begin{figure}
\centerline{\includegraphics[width=0.45\textwidth]{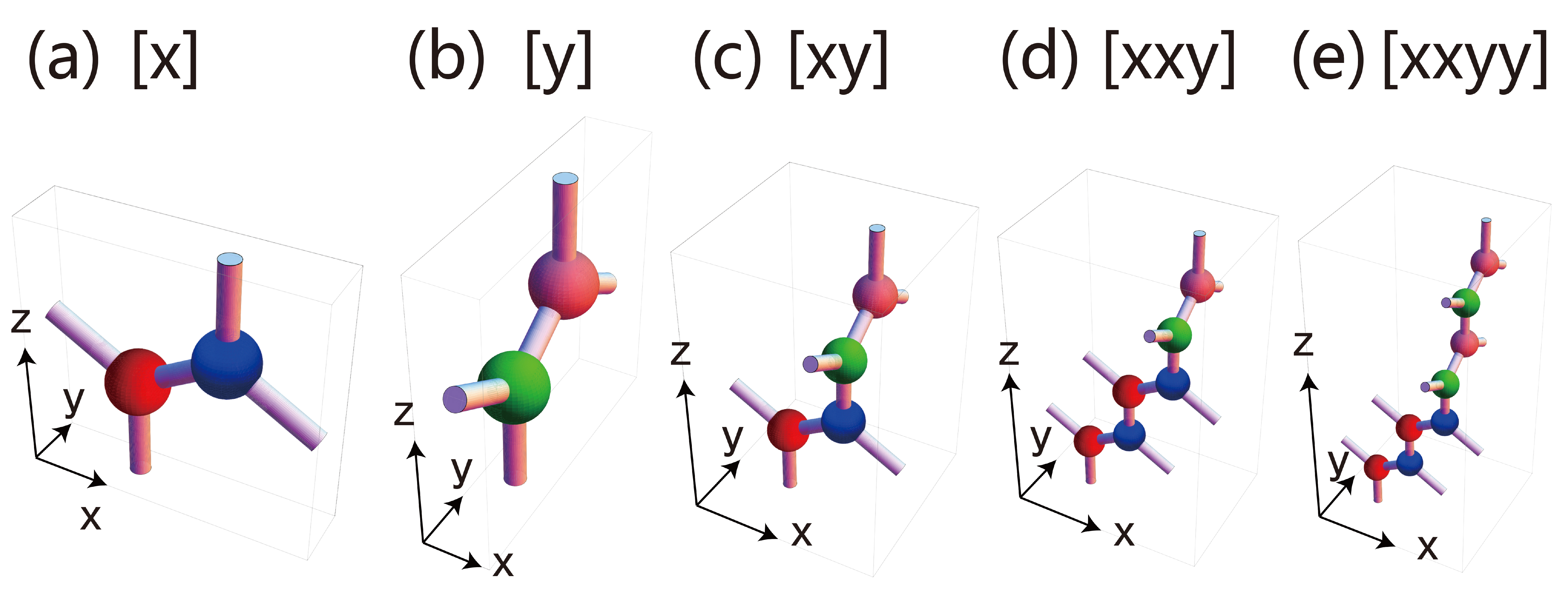}}
\caption{(Color online) Building blocks and unit cells of 3D honeycomb lattices. The building blocks are placed on the $xz$ plane (a), and $yz$ plane (b). (c)-(e) Examples of unit cells $[\alpha_1\alpha_2...\alpha_N]$ made by sewing the above blocks.
}
\label{Structure}
\end{figure}

\paragraph{Nodal ring magnon.} The 3D honeycomb lattices can be labeled by the structure units $[\alpha_1\alpha_2...\alpha_N]$\cite{Ezawa}, as shown in FIG. \ref{Structure}.  The Heisenberg model  can be explicitly written as
\begin{align}
	H & = - J \sum_{<ij>}\textbf{S}_i \cdot \textbf{S}_j,
	    \label{eq:Hamiltonian}
\end{align}
where $J$ is the isotropic exchange interaction strength, and $\textbf{S}_i$ denotes the spin on site $i$.
For linear spin-wave approximation (no magnon-magnon interactions), the Holstein-Primakoff (HP) transformation \cite{HP} reads
\begin{align}
	S^z &\rightarrow S - a^\dagger a, \\
	S^+ = S^x + \mathrm{i} S^y &\rightarrow \sqrt{2S}  a, \\
	S^- = S^x - \mathrm{i} S^y &\rightarrow \sqrt{2S}  a^\dagger,
\end{align}
where $a^\dagger$ is the magnon creation operator while $a$ the annihilation one, and they obey the boson commutation rule.
Under HP transformation and keeping quadratic terms, the nearest neighbor spin interaction terms have the general form
\begin{align}
	- J \textbf{S}_i \cdot \textbf{S}_j=-JS(-a_i^\dagger a_i-a_j^\dagger a_j+a_i^\dagger a_j+a_ia_j^\dagger).
\end{align}
Then after Fourier transformation of the Hamiltonian (\ref{eq:Hamiltonian}), we obtain its quadratic form
\begin{align}
H=\sum_k\Psi^\dagger(k)\left(
\begin{array}{cc}
A(k)&B^\dagger (k)\\
B(k)&0            \\
\end{array}
\right)\Psi(k),
\end{align}
and the basis is $\Psi^\dagger(k)=(a_{1k}^\dagger,...,a_{2N,k}^\dagger,
a_{1,-k},...,a_{2N,-k}).$

For a ferromagnetic ground state, the matrix $B(k)$ vanishes. We only need
to diagonalize $A(k)$ directly to get the spin wave spectra. The matrix $A(k)$ is given by
\begin{align}
A(k)=\left(
\begin{array}{ccccccccc}
3&f_{\alpha_1}&0&0&0&\ldots&0&0&f_z^*\\
f_{\alpha_1}^*&3&f_z&0&0&\ldots&0&0&0\\
0&f_z^*&3&f_{\alpha_2}&0&\ldots&0&0&0\\
0&0&f_{\alpha_2}^*&3&f_z&\ldots&0&0&0\\
0&0&0&f_z^*&3&\ldots&0&0&0\\
\vdots&\vdots&\vdots&\vdots&\vdots&\ddots&\vdots&\vdots&\vdots\\
0&0&0&0&0&\ldots&3&f_z&0\\
0&0&0&0&0&\ldots&f_z^*&3&f_{\alpha_N}\\
f_z&0&0&0&0&\ldots&0&f_{\alpha_N}^*&3\\
\end{array}
\right)JS,
\end{align}
where
\begin{align}
f_x&=2cos(\frac{\sqrt{3}}{2}k_x)e^{i\frac{k_z}{2}}, \\
f_y&=2cos(\frac{\sqrt{3}}{2}k_y)e^{i\frac{k_z}{2}}, \\
f_z&=e^{-ik_z}.
\end{align}
The magnon spectrum is determined by $det(A(k)-E*I)=0$.
We find that there are robust magnon nodal rings that reside in the $k_z=0$ plane with energy $3JS$.
Hence,  the position of the nodal rings are analytically determined by
$det(A(k)-3JS*I)=0$. Specifically, it is
\begin{align}
(\prod_{n=1}^Nf_{\alpha_n}-f_z^{*N})(\prod_{n=1}^Nf_{\alpha_n}^*-f_z^N)=0,
\end{align}
which leads to
\begin{align}
\prod_{n=1}^Nf_{\alpha_n}=f_z^{*N}
\end{align}
and
\begin{align}
(f_x)^{N_x}(f_y)^{N_y}=(f_z^*)^{N_x+N_y},
\end{align}
where $N_x$ ($N_y$) is the number of $x$'s ($y$'s) in $[\alpha_1\alpha_2...\alpha_N]$.
Finally, the solution is
\begin{align}
k_z=0,(2cos(\frac{\sqrt{3}}{2}k_x))^{N_x}(2cos(\frac{\sqrt{3}}{2}k_y))^{N_y}=1,
\end{align}
 which proves  that the rings are in the $k_z=0$ plane. Thus, all this class of 3D honeycomb lattices
host nodal ring magnons.

\begin{figure}
\centerline{\includegraphics[width=0.45\textwidth]{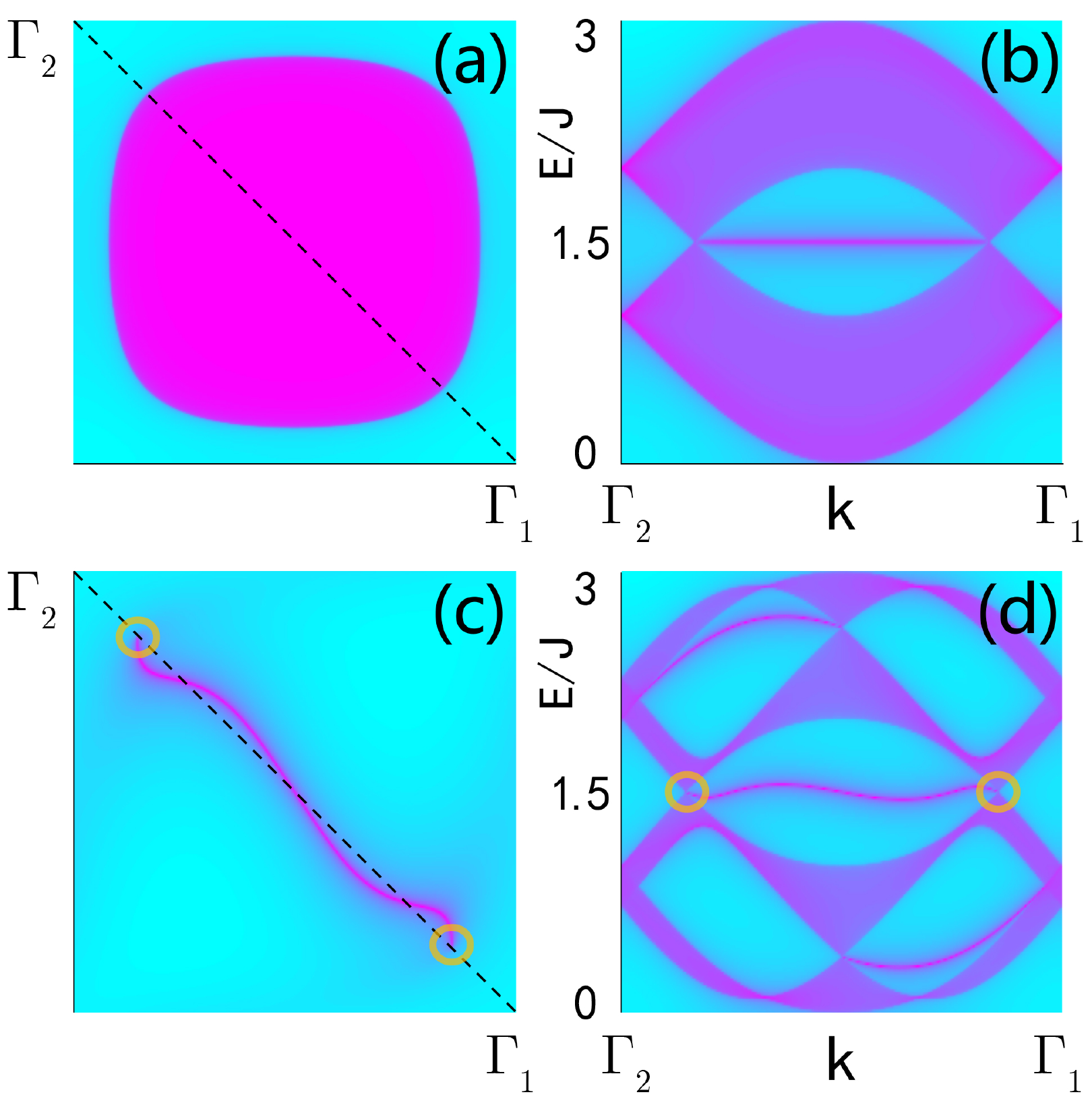}}
\caption{(Color online) Magnon surface states of the hyperhoneycomb lattice in the (001) surface. (a) The drumhead surface states at $E=3JS$
in the nodal ring phase, with the edge of the drumhead determined by the projection of the ring in the bulk.
(b) The surface states along the surface Brillouin zone path denoted by the dashed line in (a).  The
drumhead surface states are dispersionless. (c) The surface magnon arc at $E=3JS$ in the Weyl phase. (d) The surface states
along the Brillouin path denoted by the dashed line in (c). The chiral surface states are protected by the
Weyl points in the bulk. Yellow circles in (c) and (d) mark the projection of the Weyl points. Note that in the Weyl phase there are
also Weyl points and corresponding chiral surface states in the higher and lower energy than $3JS$. Here we have set $S=\frac{1}{2}$, $D=0.3J$.
}
\label{SS}
\end{figure}

To be more explicit,  we calculate the surface states of the 3D
honeycomb lattice with unit $[xy]$, which is known to be the hyperhoneycomb lattice. Due to the bulk-boundary correspondence,
there are dispersionless drumhead surface states in the (001) surface, as shown in FIG. \ref{SS} (a) and (b), which is protected by the nodal ring in the $k_z=0$ plane in the bulk.

\paragraph{Weyl magnon.} In the 2D honeycomb lattice, a proper DM interaction can gap out  Dirac magnon excitations
and drive the system into a 2D topological magnon insulator with chiral magnon edge states \cite{Owerre1}. Here we show
that in the 3D honeycomb lattices, the same kind DM interaction can gap each magnon nodal ring into two Weyl points with
opposite charges.

We consider a next-nearest neighbor DM interaction given by
\begin{align}
H_{DM}=\sum_{<<ij>>}\nu_{ij}\textbf{D}\cdot\textbf{S}_i \times \textbf{S}_j,
\end{align}
where $\nu_{ij}=\pm1$ corresponding to the next-nearest neighbor anticlockwise or clockwise path along the nearest ones with respect to the $\hat{d}=(-1,1,0)$ direction respectively. We assume the DM interaction vector and ferromagnetic spin both along  the $\hat{d}$ direction with $\textbf{D}=D\hat{d}$.
Similarly, after HP and Fourier transformations,  we get the Hamiltonian matrix for $H_{DM}$
\begin{widetext}
\begin{align}
h_{DM}=\left(
\begin{array}{ccccccccc}
-D_{\alpha_1}&0&-D_{\alpha_1}'&0&0&\ldots&0&-D_{\alpha_N}'^*&0\\
0&D_{\alpha_1}&0&D_{\alpha_2}'&0&\ldots&0&0&D_{\alpha_1}'^*\\
-D_{\alpha_1}'^*&0&-D_{\alpha_2}&0&-D_{\alpha_2}'&\ldots&0&0&0\\
0&D_{\alpha_2}'^*&0&D_{\alpha_2}&0&\ldots&0&0&0\\
0&0&-D_{\alpha_2}'^*&0&-D_{\alpha_3}&\ldots&0&0&0\\
\vdots&\vdots&\vdots&\vdots&\vdots&\ddots&\vdots&\vdots&\vdots\\
0&0&0&0&0&\ldots&D_{\alpha_{N-1}}&0&D_{\alpha_N}'\\
-D_{\alpha_N}'&0&0&0&0&\ldots&0&-D_{\alpha_N}&0\\
0&D_{\alpha_1}'&0&0&0&\ldots&D_{\alpha_N}'^*&0&D_{\alpha_N}\\
\end{array}
\right)DS,
\end{align}
\end{widetext}
where
\begin{align}
D_x&=-2sin(\sqrt{3}k_x), \\
D_y&=-2sin(\sqrt{3}k_y), \\
D_x'&=2sin(\frac{\sqrt{3}}{2}k_x)e^{i\frac{3}{2}k_z}, \\
D_y'&=2sin(\frac{\sqrt{3}}{2}k_y)e^{i\frac{3}{2}k_z}.
\end{align}
The magnon spectrum is determined by $det(A(k)+h_{DM}-E*I)=0$.
We calculate the spectrum of the lattices with units $[xy]$, $[xxyy]$
and $[xxyyy]$ numerically.  For every lattice, each nodal ring is gapped into
two Weyl points with opposite charges.  Furthermore, the positions of the Weyl points are located  in the $k_z=0$ plane  with energy $E=3JS$.

The solution is determined by
\begin{align}
det(A(k)+h_{DM}-3JS*I)=0.
\label{eqak}
\end{align}
However, it is too complicated to obtain the general analytic expression for the position
of the Weyl points. We have numerically solved Eq.\ref{eqak} in the above three cases. It is found that all nodal rings are gapped into two Weyl points. We suggest that these results should be general for all the 3D honeycomb lattices.  The generality can be further argued by the flux argument of the
DM interaction \cite{Owerre1}.

We calculate the (001) surface states of the lattice with  the unit $[xy]$.  As shown in FIG. \ref{SS},
the projections of the Weyl points are connected by the surface magnon arc, which is formed
by the chiral surface states.

\paragraph{Conclusion.} Use linear spin-wave approximation,  we study the magnon excitations
in a wide class of 3D honeycomb lattices with a ferromagnetic ground state. They all host
magnon nodal rings that lie in the same plane in the momentum space. Furthermore, the next nearest neighbor DM interaction
 can  introduce nontrivial spin chirality and gap each nodal ring into two Weyl points with
opposite charges. Our work enriches the topological magnon matter in three dimension systems.

\paragraph{Acknowledgements.} This work is supported by the Ministry of Science and Technology of China 973 program(Grant
No. 2015CB921300), National Science Foundation of China
(Grant No. NSFC-11334012), and the Strategic Priority Research Program of CAS (Grant No. XDB07000000).

\end{document}